\documentclass[10pt, a4paper,english]{article} 
\usepackage{csquotes}
\usepackage{float}
\usepackage{graphicx}

\usepackage{amsmath, amssymb, amsthm}
\usepackage{pictex, dcpic}
\usepackage{xcolor}

\def\al{\alpha}
\def\be{\beta}
\def\de{\delta}
\def\ga{\gamma}

\def\ep{\epsilon}
\def\io{\iota}

\def\ze{\zeta}
\def\om{\omega}
\def\si{\sigma}
\def\vp{\varphi}

\def\Te{\Theta}
\def\La{\Lambda}

 \def\calC{{\hbox{\cal C}}}

 \def\D{\mathbb{D}}
 \def\E{\mathbb{E}}
 \def\F{\mathbb{F}}

 \def\R{\mathbb{R}}

\def\Con{{\hbox{Con}}}

\def\diag{{\hbox{diag}}}

\def\ip{\hbox to4pt{\leaders\hrule height0.3pt\hfill}\vbox to8pt{\leaders\vrule width0.3pt\vfill}\kern 2pt}

\def\del{\partial}
\def\na{\nabla}

\def\Lie{\hbox{\LieFont \$}}

\def\arr{\rightarrow}

\def\then{\Rightarrow}

\def\barJ{\bar J}

\def\calH{{{\cal H}}}

\def\frac[#1/#2]{\hbox{$#1\over#2$}}
\def\Frac[#1/#2]{{#1\over#2}}
\def\({\left(}
\def\){\right)}
\def\[{\left[}
\def\]{\right]}
\def\^#1{{}^{#1}_{\>\cdot}}
\def\_#1{{}_{#1}^{\>\cdot}}
\def\Label=#1{{\buildrel {\hbox{\fiveSerif \ShowLabel{#1}}}\over =}}
\def\<{\kern -1pt}

\def\Lie{\pounds}

\def\Uvec#1{\vbox{\mathsurround=0pt\ialign{##\crcr
     $\scriptscriptstyle\rightharpoonup$\crcr\noalign{\kern1pt\nointerlineskip}
     $\hfil\displaystyle{#1}\hfil$\crcr}}}
\def\Dvec#1{\vbox{\mathsurround=0pt\ialign{##\crcr
     $\scriptscriptstyle\rightharpoondown$\crcr\noalign{\kern-7pt\nointerlineskip}
     $\hfil\displaystyle{#1}\hfil$\crcr}}}

\def\uvecu{\vbox{\mathsurround=0pt\ialign{##\crcr
     $\scriptscriptstyle\rightharpoonup$\crcr\noalign{\kern1pt\nointerlineskip}
     $\hfil\displaystyle{u}\hfil$\crcr}}}
\def\dvecu{\vbox{\mathsurround=0pt\ialign{##\crcr
     $\scriptscriptstyle\rightharpoondown$\crcr\noalign{\kern-7pt\nointerlineskip}
     $\hfil\displaystyle{u}\hfil$\crcr}}}

\def\uvecbe{\vbox{\mathsurround=0pt\ialign{##\crcr
     \kern3pt$\scriptscriptstyle\rightharpoonup$\crcr\noalign{\kern1pt\nointerlineskip}
     $\hfil\displaystyle{\be}\hfil$\crcr}}}
\def\dvecbe{\vbox{\mathsurround=0pt\ialign{##\crcr
     \kern1pt$\scriptscriptstyle\rightharpoondown$\crcr\noalign{\kern-10pt\nointerlineskip}
     $\hfil\displaystyle{\be}\hfil$\crcr}}}

\def\uvecn{\vbox{\mathsurround=0pt\ialign{##\crcr
     $\scriptscriptstyle\rightharpoonup$\crcr\noalign{\kern1pt\nointerlineskip}
     $\hfil\displaystyle{n}\hfil$\crcr}}}
\def\dvecn{\vbox{\mathsurround=0pt\ialign{##\crcr
     $\scriptscriptstyle\rightharpoondown$\crcr\noalign{\kern-7pt\nointerlineskip}
     $\hfil\displaystyle{n}\hfil$\crcr}}}

\def\uvecm{\vbox{\mathsurround=0pt\ialign{##\crcr
     $\scriptscriptstyle\rightharpoonup$\crcr\noalign{\kern1pt\nointerlineskip}
     $\hfil\displaystyle{m}\hfil$\crcr}}}
\def\dvecm{\vbox{\mathsurround=0pt\ialign{##\crcr
     $\scriptscriptstyle\rightharpoondown$\crcr\noalign{\kern-7pt\nointerlineskip}
     $\hfil\displaystyle{m}\hfil$\crcr}}}

\def\uvecN{\vbox{\mathsurround=0pt\ialign{##\crcr
     \kern3pt$\scriptscriptstyle\rightharpoonup$\crcr\noalign{\kern1pt\nointerlineskip}
     $\hfil\displaystyle{N}\hfil$\crcr}}}
\def\dvecN{\vbox{\mathsurround=0pt\ialign{##\crcr
     \kern0pt$\scriptscriptstyle\rightharpoondown$\crcr\noalign{\kern-10pt\nointerlineskip}
     $\hfil\displaystyle{N}\hfil$\crcr}}}

\def\uvecu{\vbox{\mathsurround=0pt\ialign{##\crcr
     $\scriptscriptstyle\rightharpoonup$\crcr\noalign{\kern1pt\nointerlineskip}
     $\hfil\displaystyle{u}\hfil$\crcr}}}
\def\dvecu{\vbox{\mathsurround=0pt\ialign{##\crcr
     $\scriptscriptstyle\rightharpoondown$\crcr\noalign{\kern-7pt\nointerlineskip}
     $\hfil\displaystyle{u}\hfil$\crcr}}}

\def\uvecw{\vbox{\mathsurround=0pt\ialign{##\crcr
     $\scriptscriptstyle\rightharpoonup$\crcr\noalign{\kern1pt\nointerlineskip}
     $\hfil\displaystyle{w}\hfil$\crcr}}}
\def\dvecw{\vbox{\mathsurround=0pt\ialign{##\crcr
     $\scriptscriptstyle\rightharpoondown$\crcr\noalign{\kern-7pt\nointerlineskip}
     $\hfil\displaystyle{w}\hfil$\crcr}}}

\def\uvecv{\vbox{\mathsurround=0pt\ialign{##\crcr
     $\scriptscriptstyle\rightharpoonup$\crcr\noalign{\kern1pt\nointerlineskip}
     $\hfil\displaystyle{v}\hfil$\crcr}}}
\def\dvecv{\vbox{\mathsurround=0pt\ialign{##\crcr
     $\scriptscriptstyle\rightharpoondown$\crcr\noalign{\kern-7pt\nointerlineskip}
     $\hfil\displaystyle{v}\hfil$\crcr}}}

\def\barJ{\kern 3pt \bar {\kern -3pt J}}

\def\ShowLabel#1{\ref{#1}}

\def\ss{\smallskip}
\def\ni{\noindent}

\def\eq#1{\begin{equation}#1\end{equation}}
\def\eqLabel#1#2{\begin{equation}#1\label{#2}\end{equation}}

\def\Cases#1{\begin{cases}#1\end{cases}}

\def\Align#1{\begin{aligned}#1\end{aligned}}

\def\eqs#1{\eq{\Align{#1}}}

\long\def\Note#1{\blockquote{\footnotesize #1}}

\date{}

\def\Figure[#1]#2{\begin{figure}[htbp] 
   \centering
   \includegraphics[#1]{#2} }

\def\EndFigure{\end{figure}}

\def\Item[#1]{\item[#1]}

\begin{document}



\title{Lecture Notes in Loop Quantum Gravity.\\
LN4: Hamiltonian framework}

\author{\small  L.Fatibene$^{a,b}$, M.Ferraris$^a$, A. Orizzonte$^{a}$\\
\\
\small$^a$ Department of Mathematics {\it``Giuseppe Peano''}, University of Torino (Italy)\\
\small$^b$ Ist. Naz. Fisica Nucleare (INFN) - Sezione Torino - Iniziativa spec. QGSKY (Italy)\\
}

\maketitle

\begin{abstract}
We discuss a covariant setting for Hamiltonian formalism in a relativistic field theory
and we use this to obtain again the properties of Hamilton principal functional
in Newtonian mechanics, relativistic mechanics, Klein-Gordon, electromagnetism, and Ashtekar-Barbero-Immirzi gravitational theory.

\end{abstract}

\section{Introduction}

Let us take a break from the specific problem of LQG and let us look back at it from a more general viewpoint.
In LN3, we discussed Hamilton-Jacobi setting for Holst formulation of GR.
The main aim there was to point out that some of the field equations are in fact constraint equations, they simply restrict
initial conditions allowed so that, after we solved the Cauchy problem in a compact bubble $D$, we are able to build out of it
a solution of the initial covariant equations; see \cite{LN2}, \cite{LN3}. 

As a matter of fact, Hamilton-Jacobi equations are properties of the Hamilton principal functional, namely the action functional $S[\si]$ computed along solutions.
In LN3, we started from a generic action functional and we showed that it is
in fact a boundary functional, it depends on fields at the boundary of the bubble, namely what we called the {\it pre-quantum configuration}.

Of course, usually this analysis is carried out in the Hamiltonian framework, where we have further structures which help the analysis,
although there is no general agreement on how to get a Hamiltonian framework for a relativistic theory.
In field theory, we have a number of, slightly different, Hamiltonian formalisms; see \cite{GIMPSY}, \cite{Rovelli1}, \cite{ADM}, \cite{Sarda}, \cite{SardaBook},
\cite{DeDonder}, \cite{Ferraris}.

The different Hamiltonian formalisms available highlight different aspects: some preserve general covariance, 
some reproduce Hamiltonian formalism for mechanics and its geometric structure.
Generally speaking, in field theories one has two different sources of problems with Hamiltonian formalism.

First of all, with more that one single independent coordinate, the physical state is parameterized by boundary fields rather than by  
numbers. That is the issue one refers when saying field theories resemble infinite dimensional mechanical systems.
Because of this issue, Hamilton principal functional in field theory is a {\it functional}.

On the other hand, dynamics in field theories are almost always degenerate, maybe only with the exception Klein-Gordon field.
Usually, one cannot solve for all derivatives of fundamental fields. The Lagrangian in GR (or in Maxwell electromagnetism) depends
on the curvature, which in turns depends only on the skew part of derivatives of the connection. 
That directly shows that one has no much hope to invert Legendre transform, as it often happens in mechanics.
Moreover, field theories typically have huge symmetry groups. 
Via hole argument, one directly sees that Cauchy problem is ill-posed, typically one has to restrict Cauchy fields to a subset which is usually dependent of the observers. As a consequence some of the field equations do not participate to the Cauchy problem and provide further constraints on initial conditions.

The analysis of constraints which comes from Legendre transform being degenerate and the ones which arise from dynamics is quite complicated in general, especially if one wants to keep the original covariance unbroken or at least under control.

As a matter of fact, it is useful to review the Hamiltonian formalism from scratch. 
Eventually, it has to be an equivalent way of formulating the dynamics. Since constraint equations are from field equations, they can be discussed in either formulation.
Although we tend to understand the difference between Lagrangian and Hamiltonian formulations of potentially the same theory in terms of the different structures they provide us (Poisson brackets, symplectic structure, \dots), it might be more effective to start from a more fundamental perspective.

As Hamilton himself claimed \cite{Hamilton}:

\Note{{\it Lagrange's function states,
Mr Hamilton's function would solve the problem. 
The one serves to form the differential equations of motion, the other would give their integrals}
}
\ss

Euler-Lagrange framework is a way of obtaining the equations, while Hamiltonian formalism is a way of characterizing the solution space 
of a theory.
In this view, the language used becomes less compelling. 
The Hamiltonian point of view focus on solution space, on physical states, it defines the phase space of a theory, regardless the mathematical framework we use.

Of course, what can be said about the solutions is also encoded in the field equations. 
What we did in LN2 and LN3 is an analysis of solutions spaces or phase space, namely a Hamiltonian  discussion although we used a Lagrangian setting for it.

In fact, in the simple case of a regular Lagrangian dynamics in  mechanics, the two viewpoints are equivalent.
Any point in $\R\times TQ$ can be taken as an initial condition for the system and, in view of the Cauchy theorem which holds true in that case,
it identifies a unique solution. The Legendre transform is one-to-one thus what can be said on $\R\times TQ$ can be equivalently moved onto $\R\times T^\ast Q$.
Either $\R\times TQ$ or $\R\times T^\ast Q$ can be regarded as the phase space of the theory, we have one solution for each point there.

On cotangent bundle, we have a natural symplectic structure, which in fact can be dragged back on the tangent bundle
(see \cite{Sutiriou}). The only difference, is that the symplectic structure which is canonical on the cotangent bundle, is not canonical on the Lagrangian setting since the Legendre map (thence the induced symplectic structure on $TQ$) depends on the Lagrangian.

When we allow non-regular dynamics, the Legendre transform becomes degenerate, for example, it defines a constraint in $S\subset T^\ast Q$.
If we use velocities in this case, different velocities are projected onto the same initial conditions in $T^\ast Q$ and ultimately correspond to a single solution.
The symmetry between tangent and cotangent is broken: physical states are points in $S\subset T^\ast Q$.

In field theory, we witness a {\it further} phenomenon: in view of the symmetries of the system, some of the equations are not evolutionary, they become constraint of allowed initial conditions. The state of the system is not in $J^1\calC$ (which in field theory takes the role of $\R\times TQ$ in mechanics).
Degenerate dynamics account for different initial conditions to produce the same solution (under-determination),
symmetries account for some initial conditions to be excluded (over-determination).
No surprise that when the two issues are present together, the situation becomes cryptic.
Over-determination restricts both $J^1\calC$ in the Lagrangian formalism and $S$ in the Hamiltonian formalism.

Here we want to discuss how to define a covariant Hamiltonian framework in a, possibly degenerate, field theory
in order to have a least a framework to compare different Hamiltonian frameworks, to compare their structures 
and maintain covariance under control.
We do not expect this analysis to add anything substantial to LN3: we still have the same constraint equations,
the same over and under-determinations. These are properties of the field equations, not of the framework we use to obtain them.

Of course, not adding anything new is not necessarily implying one cannot be easier than the other, or highlight relations with a more familiar situations, as e.g.~in mechanics.

\section{Legendre map in field theory}

Let us consider a field theory described by a Lagrangian dynamics on a configuration bundle $[C\arr M]$.
A {\it configuration} is a global section $\si:M\arr C$ of such bundle.
This whole geometric structure is encoded into the global properties of solutions, in particular in their transformation rules from one observer to another, observers being identified with trivializations of the bundle.

There is no reason or real advantage in considering local sections if we want to maintain information about transformation rules as we have to do in relativistic theories or gauge natural theories.

As the (possibly higher order) tangent bundles $[T^kQ\arr \dots \arr TQ\arr Q]$ are defined to account for total derivatives of curves and consequently they provide a definitive framework to deal with ODE on a general manifold  $Q$, the same occurs in field theory: we define jet prolongations $[J^kC\arr \dots \arr J^1C\arr C\arr M]$ which account for partial derivatives of sections, i.e.~of fields, and consequently they provide the definitive framework to deal with PDE on a general manifold and configuration bundle.

\Note{
For any fibered coordinate system $(x^\mu, y^i)$ on $C$, configurations are locally expressed as $y^i(x)$, so that $x^\mu$ are called {\it independent coordinates}, and $y^i$ are {\it field coordinates}. 
A configuration precisely gives the field values at any point in the base spacetime $M$.

Accordingly, we define fibered coordinates $j^ky:=(x^\mu, y^i, y^i_\mu, \dots, y^i_{\mu_1, \dots \mu_k})$ on the jet prolongations.
The coordinates $y^i_\mu$ are meant to represent first partial derivatives of fields, $y^i_{\mu\nu}$ are understood to be symmetric and to represent second partial derivatives, and so on.

In general, we can change fibered coordinates on $C$ as
\eq{
x'^\mu=x'^\mu(x)
\qquad
y^i = y^i(x, y)
}
and we understand that change of fibered coordinates on $C$ induces a change of coordinates on the jet prolongations, for example
\eq{
y^i_\mu = \bar J_\mu^\nu \(J^i_\nu + J^i_k y^k_\nu\)
}
where $\bar J_\mu^\nu$ is the anti Jacobian of the transformation $x'^\mu=x'^\mu(x)$, while $(J^i_\nu, J^i_k )$  are the Jacobians of $ y^i(x, y)$.
}

If we fix a vector bundle $[E\arr C\arr M]$, we define a {\it $k$-order differential operator} to be a vertical map $\D: J^kC\arr E$.
The pre-image of the zero section $0:M\arr E$ with respect to the differential operator $\D: J^kC\arr E$
defines a subset $S\subset J^kC$ which represents a PDE, just in an intrinsic and geometric way, invariant with respect to change of coordinates, on $C$.

\Note{
On $E$ we can fix coordinates $(x^\mu, y^i, e_A)$ so that the differential operator is locally expressed as
\eq{
\D: J^kC\arr E: j^ky \mapsto (x^\mu, y^i, e_A(j^ky))
}
while the associated PDE, for example for $k=1$, are
\eq{
 e_A(x^\mu, y^i(x), \del_\mu y^i(x))=0
}
}

In a variational setting, dynamics is defined out of a {\it Lagrangian} $L$.
A $k$-order Lagrangian can be considered both as a horizontal form on $J^kC$, or as a bundle map $L:J^kC\arr A_m(M)$, where $A_m(M)$
is the bundle of $m$-forms on $M$ (pulled-back on $C$ along the projection map $\pi:C\arr M$).
In both cases, the Lagrangian is in the form 
\eq{
L= L(x^\mu, y^i, y^i_\mu, \dots, y^i_{\mu_1\dots\mu_k}) d\si
}
where $d\si=dx^0\land \dots dx^{m-1}$ is the local basis of $m$-forms defined by local coordinates $x^\mu$ on $M$.

Variational calculus can be formulated in terms of maps between bundles, which are finite dimensional manifolds, with not much, if any, reference to any functional spaces. In this view, variational calculus is a way to define bundle maps, 
which eventually can be evaluated on a configuration. Evaluation along a configuration ideally has to be the last things ones do.
That is why, in mechanics, one does not really fixes a curve to find equations of motions and uses the coordinate on the tangent bundle instead.
That is why we can write the equation of a pendulum as a dynamical system on a cylinder $TS_1$, that is a submanifold of $T^2S_1\subset TTS_1$, namely as
\eq{
a+\om^2 \sin( x) =0
}

Given a compact region $D\subset M$ and a configuration $\si:M\arr C$ we can lift that to a section $j^k\si:M\arr J^kC$ of any jet prolongation,
in particular the one with the $k$ equal to the order of the Lagrangian we used to specify the dynamics.
We can define the {\it action} of the configuration $\si$ in the domain $D$ as
\eq{
A_D[\si]=\int_D (j^k\si)^\ast L
}
which is well defined ($D$ is compact and $(j^k\si)^\ast L$ is a regular form on $D$, thus the integral exists and it is finite).

We define a {\it deformation} on $D$ is a vertical vector field $X= X^i(x, y)\del_i$ on $C$,
which has a vertical flow $\Phi_s: C\arr C$ which in fact drags a section $\si$ to define a 1-parameter family of sections $\si_s:=\Phi_s\circ  \si$.

We have a bundle $[V(C)\arr C]$ for vertical vectors on $C$, with coordinates $(x^\mu, y^i, X^i)$.

It is pretty easy to show that the deformation of the action can be expressed as a bundle map
\eq{
\de L: J^kC\arr V^\ast(J^{k}C)\otimes_C A_m(M)
}

\Note{
The deformation $X$ also can be lifted to the jet prolongations, in particular $j^kX$ is a section of $[V(j^kC)\arr J^kC \arr C]$.
Then we can pair 
\eq{
\langle \de L|j^k X\rangle: J^kC\arr A_m(M)
}
which encodes for the deformation of the Lagrangian.

Let us set
\eq{
\pi_i= \Frac[\del L/\del y^i]
\qquad
\pi_i^\mu= \Frac[\del L/\del y^i_\mu]
\qquad
\pi_i^{\mu\nu}= \Frac[\del L/\del y^i_{\mu\nu}]
\qquad
\dots
}
For a first order Lagrangian, we have
\eq{
\langle \de L|j^1 X\rangle:J^1C\arr A_m(M): (j^1y)\mapsto  \(\pi_i X^i+ \pi_i^\mu d_\mu X^i\) d\si
}
}

Variational calculus can be encoded in this way by global maps between bundles, namely
\eqs{
& \E: J^{2k}C\arr V^\ast(C)\otimes_C A_m(M) \cr
&  \F: J^{2k-1}C\arr V^\ast(J^{k-1}C)\otimes_C A_{m-1}(M) \cr
} 
so that one has the so-called {\it first variation formula}
\eq{
\langle \de L|j^k X\rangle = \langle \E|X\rangle + d \langle \F|j^{k-1} X\rangle
}
which is satisfied by any deformation as equality of maps $J^kC\arr A_m(M)$.

\Note{
That can be easily shown by choosing coordinates and locally going through the usual integrations by parts one does in variational calculus.
For globality, one can fix the order $k$ and check transformation laws.
One can see the proof in \cite{book1}, for order $k=1,2$ in \cite{book2}.

For a first order Lagrangian, we have 
\eq{
 \E= (\pi_i -d_\mu \pi_i^\mu)\> \bar dy^i\otimes d\si
\qquad
 \F=  \pi_i^\mu\> \bar dy^i\otimes d\si_\mu
}
where $\bar dy^i$ is a the dual basis of $\del_i$ in $V(C)$.

The whole first variation formula can be evaluated along a configuration $\si$, namely setting $y^i=y^i(x)$, to become the usual local computation one does for variational calculus.
}

Notice that $V^\ast (C)\otimes_C A_m(M)$ is a vector bundle therefore, by definition, $\E$ is a differential operators of order (at most) $2k$,
the corresponding PDE, $(\E_i \circ j^{2k}\si)=0$, are the {\it Euler-Lagrange equations}, the {\it field equations} of our system.

\subsection{The Legendre map and phase bundle}

Let us restrict, for the sake of simplicity, to first order Lagrangians.
We define the {\it Legendre map} as
\eq{
\F: J^1 C\arr V^\ast (C) \otimes A_{m-1}(M)
:j^1y \mapsto  \pi_i^\mu (j^1y)\> \bar d y^i \otimes d\si_\mu
}

This defines the {\it (Hamiltonian) phase bundle} $P(C)= V^\ast (C) \otimes A_{m-1}(M)$, namely
the bundle in which momenta $\pi_i^\mu$ live. 
It has coordinates $(x^\mu, y^i,p_i^\mu)$ and as usual that transformations on $[C\arr M]$ induce transformations on $P(C)$
\eq{
 \Cases{
   x'^\mu= x'^\mu(x)\cr
   y'^i = y'^i(x, y)\cr
 }
 \qquad\qquad\then\quad
   p'^\mu_i = \bar J J^\mu_\nu p^\nu_k \bar J^k_i
}

In other words, we have defined a functor $P(\cdot)$ 
which associates to a configuration bundle $C$ its Hamiltonian phase bundle $P(C)$.
 
Of course, in field theory, we do not require $\F$ to be invertible since this does not happen in most models.
However, let us require as regularity that the map $\F$ is constant rank, i.e.~the image $\La= \F(J^1C) \subset P(C)$  is a sub-manifold
which is called the {\it primary constraint}.

\subsection{The multisymplectic bundle and the canonical Liouville form}

Let us define {\it Poincar\'e-Cartan forms} to be of $m$-forms $\om$ on $C$ such that for any 2 vertical vectors $u, v\in V(C)$ one has
$v\ip u\ip \om =0$.

That means that locally a Poincar\'e-Cartan form $\om$ is written as
\eq{
\om = p \> d\si + p_i^\mu \>dy^i \land d\si_\mu
}
\Note{
If we change coordinates in $C$ that induced transformations of Poincar\'e-Cartan forms
\eqs{
\om =& p' \> d\si' + p'^\mu_i \>dy'^i \land d\si'_\mu=\cr
=& J p' \> d\si + J \bar J_\mu^\nu p'^\mu_i \( J^i_\al dx^\al +J^i_k \>dy^k\) \land d\si_\nu=\cr
=& J \( p' +  \bar J_\mu^\nu J^i_\nu  p'^\mu_i  \)  d\si + J \bar J_\mu^\nu p'^\mu_i J^i_k \>dy^k \land d\si_\nu
}
i.e.~transformation laws for coefficients
\eq{
p = J \( p' +  \bar J_\mu^\nu J^i_\nu  p'^\mu_i  \)
\qquad\qquad
 p_k^\nu = J \bar J_\mu^\nu p'^\mu_i J^i_k
}
}
We can define the {\it multisymplectic bundle} $Z(C)\subset A_m(C)$ as the subbundle with coordinates $(x^\mu, y^i, p, p_i^\mu)$ and transition functions above.

As it happens on cotangent bundles, on $Z(C)$ one can define a canonical {\it  Liouville $m$-form}
\eq{
\Te= p \> d\si + p_i^\mu \> dy^i\land d\si_\mu 
}
\Note{
The canonical Liouville $m$-form is a global form exactly in view of transformation laws
\eqs{
\Te=& p' \> d\si' + p'^\mu_i \> dy'^i\land d\si'_\mu =\cr
=& J p' \> d\si + J \bar J_\mu^\nu p'^\mu_i \> \( J^i_\al dx^\al + J^i_k dy^k\)\land d\si_\nu =\cr
=&J \( p'  +   \bar J_\mu^\nu p'^\mu_i J^i_\nu \) d\si + J \bar J_\mu^\nu p'^\mu_i J^i_k \> dy^k\land d\si_\nu=\cr
=&p \> d\si + p_k^\nu \> dy^k\land d\si_\nu = \Te
}

A manifold $(M, \om)$ with a closed, non-degenerare, $k$-form is called a {\it multisymplectic manifold} of degree $k$.
On the bundle $Z(C)$ we have the $(m+1)$-form $\om=d\Te$ which is clearly closed (in fact exact) and non-degenerate since
\eqs{
\vec v\ip d\Te=&  \vec v\ip \(dp \land d\si + dp_i^\mu \land dy^i\land d\si_\mu \) =\cr
=& v d\si - v^\al dp\land d\si_\al + v_i^\mu dy^i \land d\si_\mu +\cr
&\quad
- v^i dp_i^\mu \land d\si_\mu + v^\al dp_i^\mu \land dy^i\land d\si_{\mu\al}=0\cr
\then&\quad
v=0, \> v^\al=0, \> v^i=0, \> v_i^\mu =0
\quad\then\quad \vec v=0
}
Therefore $(Z(C), \om)$ is a multisymplectic bundle of degree $k=m+1$, which reduces to the usual symplectic structure in mechanics when $m=1$.
}

Also $Z(\cdot)$ is a functor which defines the multisymplectic bundle $Z(C)$ out of the configuration bundle $C$. 
Of course, we have a projection map $p: Z(C)\arr P(C)$
which is well defined exactly because $p_i^\mu$ transform in the same way on $P(C)$ and $Z(C)$.

\subsection{Poincar\'e-Cartan form in the Lagrangian setting}

The dynamics of the theory is also described by the {\it Poincar\'e-Cartan form}  on $J^1C$
\eq{
\Te_L = L(j^1y) d\si + \pi_i^\mu(j^1y)  \( d y^i - y^i_\al dx^\al\)\land d\si_\mu
= -\calH d\si + \pi_i^\mu  d y^i \land d\si_\mu 
}
where we set $\calH(j^1y) :=  \pi_i^\mu y^i_\mu -L$ for the {\it total energy}.
The  Poincar\'e-Cartan form $\Te_L $ is a global form of the type we used to define $Z(C)$, which means that the Lagrangian in fact defines a global map
\eq{
\Phi_L: J^1C \arr Z(C): j^1y \mapsto \(x, y, p=-\calH , p_i^\mu =  \pi_i^\mu \)
}
\Note{
The map is well defined since the total energy $\calH$ transforms as $-p$ in fact
\eqs{
-\calH=&-\pi_i^\mu y^i_\mu +L
=-J \( \bar J^\mu_\be \pi'^\be_k J^k_i y^i_\mu -L'\)
=-J \( \pi'^\be_k (y'^k_\be - \bar J_\be^\mu J^k_\mu)   -L'\)=\cr
=&J \( -\calH'  +  \bar J_\be^\mu J^k_\mu  \pi'^\be_k  \)
}

One can easily get that $\Te_L$ is an alternative way of describing the dynamics, on equal footing of the Lagrangian.
In fact, the action can be written as
\eq{
 A_D[\si] = \int_D (j^1\si)^\ast L
 = \int_D (j^1\si)^\ast \Te_L
}
since we have that $\om^i :=d y^i - y^i_\al dx^\al$ is a {\it contact 1-form}, i.e.
\eq{
  (j^1\si)^\ast \om^i = \( d_\al y^i  - y^i_\al \) dx^\al =0
  \qquad
   \om'^i =  J^i_k \( d y^k -y^k_\be dx^\be\) = J^i_k \om^k
}
In view of these transformation laws, we have
\eq{
 \Te'_{L'}=
 L' d\si' + \pi'^\mu_i \om'^i \land d\si'_\mu
 = J L'  d\si + J \bar J_\mu^\nu \pi'^\mu_i J^i_k \> \om^k \land  d\si_\nu
  = \Te_L
}
which implies the transformation laws for the Lagrangian $JL' = L$ (which is equivalent to the Lagrangian globality) and it, in turns, implies that of momenta
\eq{
J  \bar J_\al^\mu  \pi'^\al_k J^k_i  = \pi^\mu_i 
}
The variation of the action is then
\eqs{
\de A_D[\si] =& \int_D \Frac[d/ds] (j^1\si)^\ast \Phi_s^\ast \Te_L = \int_D  (j^1\si)^\ast \Lie_{\hat X} \Te_L=\cr
=& \int_D  (j^1\si)^\ast \(  \hat X\ip d \Te_L + d (\hat X\ip  \Te_L) \)
}
where $\hat X$ is the prolongation of the deformation to the Lagrangian phase space.
Using Stokes theorem we have
\eq{
\de A_D[\si] 
= \int_D  (j^1\si)^\ast   \hat X\ip d \Te_L +  \int_{\del D } (\hat X\ip  \Te_L) 
}
The boundary term vanishes for deformations fixed on the boundary and field equations reads as
\eq{
(j^1\si)^\ast   \hat X\ip d \Te_L=0
}
Of course, these are equivalent to the Euler-Lagrange equations.
}

Let us remark that the Poincar\'e-Cartan form $\Te_L=\Phi_L^\ast \Te$ is pull-back of the canonical Liouville form on $Z(C)$.
Also, the map $\Phi_L$ factorizes through $\F:J^1C\arr \La$, i.e.~there exists a map $\Phi_H:\La\arr Z(B)$ such that $\Phi_H\circ \F= \Phi_L$.

\Note{
\ni{\bf Proof}: The map $\F:J^1B\arr \La$ is surjective and of constant (maximal) rank. It is then a projection and it defines a fibration $\F: J^1B\arr \La$.
We need to prove that $\Phi_L$ is constant along the fibers of such a fibration.

Let us consider a curve $\ga:\R \arr J^1B: s\mapsto (x^\mu, y^i, y^i_\mu(s))$ in the fiber, then
\eqLabel{
\Frac[d/ds]\( \F\circ \ga\) = \Frac[\del L / \del y^i_\mu \del y^k_\al]  v^k_\al = 0
}{tangentVector} 
i.e.~$\F\circ \ga$ is constant.

A vector $v= v^\mu\del_\mu+v^i\del_i + v^i_\mu \del_i^\mu$ is tangent to the fiber  iff $v^\mu=0$, $v^i=0$, and (\ShowLabel{tangentVector}) is satisfied.

Now we can compute $\Frac[d/ds]\( \Phi_L \circ \ga\)$ which corresponds to compute
\eqs{
&\Frac[d/ds] p_i^\mu 
= \Frac[d/ds] \Frac[\del L / \del y^i_\mu]
= \Frac[\del L / \del y^i_\mu \del y^k_\al]  v^k_\al
=0\cr
-&\Frac[d/ds] \calH 
= \Frac[d/ds] (L- p_i^\mu y^i_\mu)
=
-\( \Frac[d/ds] p_i^\mu\)  y^i_\mu
=0
\cr
}

Since the map $\Phi_L$ is constant along the fibers, we can set $\Phi_H(x) = \Phi_L(p)$ for any $p$ such that $\F(p)=x$,
by which the map $\Phi_H$ is well defined.
}

Of course, we can set $\Te_H:= (\Phi_H)^\ast \Te$ on $\La$ and we have for free that
\eq{
\Te_L
= (\Phi_L)^\ast \Te
= (\Phi_H\circ \F)^\ast \Te
= (\F)^\ast \circ (\Phi_H)^\ast \Te
=(\F)^\ast \Te_H
}

Hence, we have this situation
\eq{
\begindc{\commdiag}[8]
\obj(162,198)[Z]{$Z(C)$}
\obj(162,135)[P]{$P(C)$}
\obj(90,135)[La]{$\La$}
\obj(20,135)[J]{$J^1C$}
\obj(18,90)[B]{$C$}
\obj(18,40)[M]{$M$}
\obj(90,90)[B1]{$C$}
\obj(90,40)[M1]{$M$}
\obj(162,90)[B2]{$C$}
\obj(162,40)[M2]{$M$}
%
%
\mor{B}{M}{}
\mor{J}{B}{}
\mor{B1}{M1}{}
\mor{La}{B1}{}
\mor{B2}{M2}{}
\mor{P}{B2}{}
\mor{J}{La}{$\F$}
\mor{La}{P}{}[\atright, \injectionarrow]
\mor{Z}{P}{}
\mor{J}{Z}{$\Phi_L$}
\mor{La}{Z}{$\Phi_H$}[\atright,\solidarrow]
\mor{M}{M1}{}[\atleft,\solidline]
\mor{M1}{M2}{}[\atleft,\solidline]
\mor{B}{B1}{}[\atleft,\solidline]
\mor{B1}{B2}{}[\atleft,\solidline]
\mor(90,93)(162,93){}[\atleft,\solidline]
\mor(90,43)(162,43){}[\atleft,\solidline]
\mor(18,93)(90,93){}[\atleft,\solidline]
\mor(18,43)(90,43){}[\atleft,\solidline]
\enddc
}

Let us fix coordinates $(x^\mu, y^i, q^a)$ on $\La$ so that the canonical embedding $\io:\La\arr P(C)$ is given by
$p_i^\mu= p_i^\mu(x, t, q)$.

If we consider a section $\tilde \si:M\arr \La: x\mapsto (x, y(x), q(x))$ of the bundle $[\La\arr C\arr M]$.
That projects on a section $\si:M\arr C$ and lift to $j^1\si:M\arr J^1C$ so that $\F\circ j^1\si= \tilde \si$.
We can express the action as 
\eq{
A_D[\tilde \si]= \int_D (j^1\si)^\ast  \Te_L
= \int_D (j^1\si)^\ast \circ (\F)^\ast \Te_H
=\int_D (\tilde \si)^\ast \Te_H
}
therefore, the form $\Te_H$ provides an equivalent formulation of dynamics.

Let us define a {\it Hamiltonian section}  to be a section $\tilde \si: M\arr \La$ iff  for any vector field $\Xi$ tangent to $\La$, one has
\eq{
(\tilde \si)^\ast \Xi \ip d\Te_H=0
}
namely, iff it is a critical point of the action.

The good thing of regarding variational calculus as a mathematical tool is that it can be nested and bent without worrying too much about physical meaning, which can be discussed later.
As a matter of fact we can rewrite the action
\eq{
A_D[\tilde \si]=\int_D (\tilde \si)^\ast \Te_H
= \int_D( p_i^\mu(x, y, q) d_\mu y^i - H(x, y, q)) d\si
}
i.e.~we can define a {\it Helmholtz  Lagrangian} on $J^1(\La)$
\eq{
L_H(j^y, q) = ( p_i^\mu(x, y, q) d_\mu y^i - H(x, y,q)) d\si
}
which is zero order in $q^a$ and first order in $y^i$.
Here the Hamiltonian is defined so that 
\eqLabel{
H(x, y, q(j^1y))= \calH(j^1y) = \pi_i^\mu(j^1y) y^i_\mu - L(j^1y)
}{Hamiltonian} 
is induced by the Lagrangian and $q(j^1y)$ is the expression of the Legendre transform $\F$.

Notice how this is slightly more convoluted that what one usually do in mechanics, all to go along with the fact that $\F:J^1C\arr \La$ may not be invertible.
Anyway, defining the Helmoltz Lagrangian $L_H$ we are back into a Lagrangian framework and, 
in fact, Euler-Lagrange equations for $L_H$ are
\eq{
\Frac[\del H/\del q^a] = \Frac[\del \pi_i^\mu / \del q^a] d_\mu y^i 
\qquad\qquad
d_\mu  \pi_k^\mu = \Frac[\del \pi_i^\mu /\del y^k]d_\mu y^i  - \Frac[\del H/\del y^k]  
}
These are the {\it Hamilton equations} for $H(x, y, q)$.
They are, {\it by construction}, equivalent to the original Euler-Lagrange equations for $L$, just as they determine critical configuration of the same (or rather equivalent) action.

\Note{
One can, maybe should, also show the equivalence explicitly. 
In case, the derivatives of (\ShowLabel{Hamiltonian}) provides identities among Jacobians which does the job.
}

\section{Mechanics}

Instead of showing equivalence in the general case we think it is more instructive to consider different cases, with different structures of under and over-determination and dynamical non-regularities.

If we downgrade this formalism to mechanics, we start from a configuration bundle $[C\arr \R]$, we set $C= \R\times Q$ where $Q$ is the configuration space of the holonomic system.
Configurations are sections of this bundle, which correspond to parameterised curves on $Q$. 

The Lagrangian phase bundle is isomorphic to $J^1C\simeq \R\times TQ$,
the Hamiltonian phase bundle is isomorphic to $P(C)\simeq \R\times T^\ast Q$,
the multisymplectic bundle is isomorphic to $Z(C)\simeq T^\ast(\R\times Q)$.
Multisymplectic bundle $Z(C)$ is a cotangent bundle, it therefore has a canonical Liouville form $\Te= p dt + p_i dq^i$ 
which defines a symplectic form $\om=d\Te$, hence a symplectic structure $(Z(C), \om)$.

Let us first consider a Newtonian like dynamics, encoded by some usual non-degenerate Lagrangian $L(t, q, u)$.
We  define the {\it momenta} as $\pi_i(t, q, u)=\Frac[\del L/\del u^i]$, we assume we can solve for velocities $u^i=u^i(t, q, p)$.

\Note{
Being $u^i(t, q, p)$ obtained by solving momenta, we have the identity $\pi_i(t, q, u(t, q, p))=p_i$ which, by derivation, gives us the identities
\eq{
\Frac[\del \pi_i/ \del q^k] + \Frac[\del \pi_i/ \del u^k] \Frac[\del u^k/\del q^j]= 0
\qquad\qquad
\Frac[\del \pi_i/ \del u^k] \Frac[\del u^k/\del p_j]= \de^j_i
} 
}

The {\it total energy} and the corresponding {\it Hamiltonian}  are set to
\eqs{
&\calH(t, q, u) = \pi_i (t, q, u)  \> u^i - L(t, q, u) 
\cr
&H(t, q, p) = p_i u^i(t, q, p) -L(t, q, u(t, q, p))
}

In this case, we have that $\La\equiv P(C)$. The relevant maps into $Z(C)$ are
\eq{
\Phi_L(j^1y) = (t, q, p= -\calH, p_i=\pi_i)
\qquad\qquad
\Phi_H(j^1y) = (t, q, p_i, p=-H)
}
from with the forms readily follows
\eq{
\Te_L= Ldt + \pi_i \om^i
\qquad\qquad
\Te_H = -H dt + p_i dq^i
}
Hamilton equations are simply
\eq{
\dot q^i = \Frac[\del H/\del p_i](t, q, p)
\qquad\qquad
\dot p_i = -\Frac[\del H/\del q^i](t, q, p)
}
The first is the inverse Legendre transform, the second is equivalent to Euler-Lagrange equation. 
Hamilton equations define a dynamical system on $\R\times T^\ast Q$ given by the Hamiltonian field 
\eq{
X_H= \del_t + \Frac[\del H/\del p_i] \Frac[\del/\del q^i] -  \Frac[\del H/\del q^i] \Frac[\del/\del p_i]
} 
which can be mapped back into $\R\times TQ$ equivalently into
\eqs{
&X_L= \del_t + u^i \Frac[\del/\del q^i] + (X_L)^i \Frac[\del/\del u^i]\cr
&\quad (X_L)^i  = A^{ij} \( \Frac[\del L/ \del q^j]  - \Frac[\del L/\del t \del u^j] -  \Frac[\del L/\del q^k \del u^j] u^k\)
} 
where $A^{ij}$ is the inverse Hessian of the Lagrangian with respect to velocities.

\Note{
Let us stress, neither the Lagrangian or the Hamiltonian are in fact {\it scalar functions}, they both are components of forms which transform accordingly. Only by restricting to special transformations, e.g.~$(t'=t+t_0, q'= q(t, q))$, they behave as functions.

Also in both frameworks, dynamics is expressed as a dynamical system, as for any normal ODE.
Moreover, the dynamical systems are mapped one into the other by Legendre transform, so that also the solutions are mapped one into the others.
}

This is in a nutshell the ordinary situation in classical mechanics.
However, we can also consider a {\it relativistic point}, which is described somehow in a similar way on $C=\R\times M$ with coordinates $(s, x^\mu)$
where $s$ is a parameter (not necessarily time)  and $M$ is spacetime (including in some sense time).
On spacetime, we are interested in {\it trajectories} (somehow {\it unparameterized} curves), thus when we specify dynamics we want it to be invariant with respect to re-parameterizations. This can easily be done by fixing the Lagrangian to be
\eq{
L = \sqrt{-g_{\mu\nu}(x)u^\mu u^\nu} + A_\mu(x)u^\mu
}
for some covector $A=A_\mu dx^\mu$. Let us fix $(M, g)$ to be Minkowski  and choose $x^\mu$ to be orthonormal coordinates so that $g_{\mu\nu}=\eta_{\mu\nu}$.

One can check that this is invariant with respect to parameterizations as we required. 
We define the momenta
\eq{
\pi_\mu := \Frac[-\eta_{\mu\nu} u^\nu/  \sqrt{-\eta_{\al\be}u^\al u^\be} ] + A_\mu
\quad\then
(\pi-A)\cdot (\pi-A) = -1  
}
The Legendre map $\F:\R\times TM\arr \R\times T^\ast M$ is clearly not surjective since $p-A$ needs to be on surface, which is our primary constraint $\La\subset  \R\times T^\ast M$.
We can use $p_i$ to parameterize $\La$ and set the embedding giving $p_0$ as a function of $p_i$.

In this case we have 
\eq{
\calH= \pi_\mu u^\mu -L= \Frac[-u\cdot u/\sqrt{-u\cdot u}]  +A\cdot u - \sqrt{-u\cdot u} - A\cdot u\equiv 0
}
thus we have
\eq{
\Te_L =  \pi_\mu dx^\mu = \Frac[-u\cdot dx/  \sqrt{-u\cdot u} ] + A\cdot dx
\qquad\qquad
\Te_H = p_\mu dx^\mu
}
and the equations turn out to be
\eq{
\dot x=u
\qquad\qquad
\dot u = u \frac[d/ds] \( \ln\(\sqrt{-u\cdot u}\)\)+ (\sqrt{-u\cdot u}) u\ip F
}
where we set $F=dA$ for the 2-form $F$.
Here reparameterizations work exactly as for the hole argument (they are symmetries and they can be compact supported in $\R$) and they  spoil uniqueness of solutions, Cauchy theorem cannot hold since when we have a solution any of its reparameterization is again a solution.

We have to discuss whether we have constraint equations as well.
In order to do that, we can change coordinates $\om^2 =-u\cdot u$ and $v^i= \Frac[u^i/u^0]$, which can be inverted as
\eq{
u^0= \Frac[\om/ \sqrt{1-  |\vec v |^2} ] 
\qquad\qquad
u^i= \Frac[\om v^i/ \sqrt{1-  |\vec v |^2} ] 
}
By obvious reasoning, in dimension 4, we can set $F_{0i}= E_i$ and $F_{ij}= \ep_{ij}{}^k B_k$, so that 
$(u\ip F)\^i=u^0 (E-  v \times B)$ and $(u\ip F)^0= u^0  v\cdot  E$, and split the second equation as
\eq{
\frac[\dot u^0/u^0] = \frac[d/ds]\(\ln(\om)\) +\om v\cdot E 
\qquad
\frac[\dot u^i/u^0] = v^i  \frac[d/ds]\(\ln(\om)\) + \om(E- v\times B)
}
These can be recast as 
\eq{
\dot \om =\dot \om
\qquad
\dot v = \om\(E-  v \times B\) - \om (v\cdot E) v
\quad\(\then\> v\cdot \dot v = \om (1-|v|^2) v\cdot E \)
} 
The first equation of these is identically satisfied, so that $\om$ is uncostrained.
For the second equation we can solve $v\cdot E = \Frac[v\cdot \dot v/ \om (1-|v|^2) ]$ and substitute it back
\eq{
\dot v +  \Frac[v\cdot \dot v/  (1-|v|^2)]  v = \om\(E-  v \times B\) 
}

\Note{
We can further simplify it. Let us fix $\ga= \frac[1/\sqrt{1-|v|^2}]$ and consider its derivative $\dot \ga = \ga^3 v\cdot \dot v$ and then compute directly
\eq{
\frac[1/\ga] \frac[d/ds]\(\ga v\) = \dot v + \frac[(v\cdot \dot v) v / 1-|v|^2] = \om\(E-  v \times B\) 
}
}

Then the $m$ equations of motions are in fact equivalent to $m-1$
\eq{
 \frac[d/ds]\(\ga v\) = \om\ga\(E-  v \times B\) 
}
which determines the physical velocity $v$ and leave $\om$ completely undetermined.
There is no constraint on initial conditions, in fact there should be $\calH=0$ which, however, is identically satisfied since $\calH\equiv 0$.

\section{Klein-Gordon  field}

The first example in field theory is (real) Klein-Gordon field, for the sake of simplicity on Minkowski spacetime.
The configuration bundle is $[C\arr M]$ with $C=M\times \R$ which is trivial so that the sections are {\it scalar} fields.
We pick coordinates $(x^\mu, \vp)$ on $C$, with $x^\mu$ orthonormal with respect to the Minkowski metric $\eta=\diag(-1, 1, 1, 1)$.

The dynamics is induced by the Lagrangian
\eq{
L= -\frac[1/2](\vp_\mu \eta^{\mu\nu} \vp_\nu + \mu^2 \vp^2)d\si
}

We define momenta as $\pi^\mu = -\vp^\mu$. The Legendre transform is trivially invertible, hence no primary constraint.
The total energy  reads as
\eq{
\calH = -\vp^\mu \vp_\mu + \frac[1/2](\vp_\mu \eta^{\mu\nu} \vp_\nu + \mu^2 \vp^2)
= -\frac[1/2](\vp_\mu \eta^{\mu\nu} \vp_\nu - \mu^2 \vp^2)
}
hence we have the Hamiltonian in the form
\eq{
H(x, \vp, p)=  -\frac[1/2](p^\mu \eta_{\mu\nu} p^\nu - \mu^2 \vp^2)
}

We can define the maps
\eqs{
\Phi_L(j^1y) =& (x, \vp, p^\mu = -\vp^\mu, p= -\calH)\cr
\Phi_H(x^\mu, \vp, p^\mu) =& (x^\mu, \vp, p^\mu, p=-H(x, \vp, p^\mu))
}
so that we get the Poincar\'e-Cartan forms
\eq{
\Te_L = Ld\si -\vp^\mu \om\land d\si_\mu
\qquad\qquad
\Te_H = -H d\si +p^\mu d\vp\land d\si_\mu
}
This is pretty much the same as regular mechanics, no constraints.
The field equations
\eq{
\eta^{\mu\nu} \vp_{\mu\nu} -\mu^2 \vp=0
}
defines a well-posed Cauchy problem.

The Hamilton principal functional is 
\eqs{
\de S =& \Frac[d/ds ]\int_{D_s} (j^1\vp_s)^\ast \Te_L = \int_{D} (j^1\vp)^\ast  \Lie_{j^1 X} \Te_L=\cr
=& \int_{D} (j^1\vp)^\ast  \(j^1 X\ip d \Te_L+ d ( X\ip  \Te_L)\)=\cr
=&\int_{D} (j^1\vp)^\ast  (j^1X) \ip d \Te_L+ \int_{\del D} (j^1\vp)^\ast  (X\ip \Te_L)
}
On solutions, the first integral vanishes, thus
\eqs{
\de S =& \int_{\del D} (j^1\vp)^\ast  (X\ip \Te_L)=\cr
=&  \int_{\del D}\(   u_\al   (  \vp^\al \vp_\be  -\frac[1/2](\vp_\mu \vp^\mu + \mu^2 \vp^2) \de^\al_\be )\de x^\be  + u_\al p^\al  \de \vp \) d\ze
}
where $d\ze$ is the standard $(m-1)$-form induced by coordinates on the boundary $\del D$.
Then, we have
\eq{
\frac[\de S/\de x^\be] =  u_\al   (  \vp^\al \vp_\be  -\frac[1/2](\vp_\mu \vp^\mu + \mu^2 \vp^2) \de^\al_\be ) =u_{\al}  \hat T^\al_\be  
\qquad
\frac[\de S/\de \vp] = u_\al p^\al  
}
which, in fact, coincide with what we found previously.

As a matter of fact, Poincar\'e-Cartan forms, either $\Te_L$ or $\Te_H$, provide us with a way of computing the Hamilton principal functional
directly. 
The relation with Poincar\'e-Cartan form also suggests a direct link with conservation laws. 
On the other hand, the relation with multisymplectic structure also opens to a definition of Poisson or symplectic structures.
We do not discuss neither here.

\section{Electromagnetic field}

We can consider electromagnetism on Minkowski spacetime $(M, \eta)$. We fix a structure $U(1)$-principal bundle $[P\arr M]$ (which, in this case, is necessarily trivial, since the base $M$ is contractible; see \cite{Steenrod}).
The electromagnetic field is described in terms of a connection on $P$ which is locally described in terms of a covariant potential $A=A_\mu dx^\mu$.

\Note{
Since $P$ is trivial, connections allow global potential, thus the space $\Con(P)$ of connections $A$ on the structure bundle $P$ is isomorphic to the cotangent bundle $T^\ast M$.
However, if it were a covector $A_\mu$ would transform as $A'_\mu = \barJ_\mu^\nu A_\nu$ while, in view of gauge transformations, in fact it transforms as $A'_\mu = \barJ_\mu^\nu (A_\nu + \del_\nu \al)$ namely, as a connection on $P$.

The isomorphism $\Con(P)\arr T^\ast M$ is not canonical, we have one isomorphism for each (global) trivialization of $P$.
Again, bundles are there to take care of transformation laws.
}

For the sake of simplicity, we restrict to orthonormal coordinates $x^\mu$, which exist on Minkowski spacetime.
Let us set $F=dA$ for the {\it field strength} and the Lagrangian can be written as
\eq{
L= -\frac[1/4] F^{\mu\nu} F_{\mu\nu}
}

We define momenta as $\pi^{\mu\nu} = -F^{\nu\mu}= F^{\mu\nu}$. 
The Legendre transform is not invertible, since of course we cannot solve for it $d_\al A_\be$ as the momenta depends only on the skew part namely, on $F_{\al\be}$.
The total energy  reads as
\eq{
\calH = F^{\mu\nu} d_\nu A_\mu - L
=  \frac[1/2]F^{\mu\nu} (d_\nu A_\mu - d_\mu A_\nu) +\frac[1/4] F^{\mu\nu} F_{\mu\nu}
= - \frac[1/4]F^{\mu\nu} F_{\mu\nu} 
}
hence we have the Hamiltonian in the form
\eq{
H(x, A, p)=  -\frac[1/4]p^{\mu\nu}p_{\mu\nu} 
}

We can define the maps
\eq{
\Phi_L(j^1y) = (p^{\mu\nu} = F^{\mu\nu}, p= -\calH)
\qquad
\Phi_H( p^{\mu\nu}) = (p^{\mu\nu}, p=-H(x, A, p^\mu))
}
so that we get the Poincar\'e-Cartan forms
\eq{
\Te_L =  \frac[1/4]F^{\mu\nu} F_{\mu\nu} d\si -F^{\mu\nu} dA_\nu \land d\si_\mu 
\qquad\qquad
\Te_H = -H d\si +p^{\nu\mu} dA_\nu \land d\si_\mu
}

The Hamilton functional reads as
\eqs{
\de S =& \int_{\del D} (j^1\si)^\ast  (X\ip \Te_L)=\cr
=&  \int_{\del D}  -u_\al ( \frac[1/4]F^{\mu\nu} F_{\mu\nu} \de_\be^\al  \de x^\be  -  F^{\al\nu} (\de x^\be d_\be A_\nu  -\de A_\nu ))   d\ze
}
Let us remark that not all fields $X= \de x^\be\frac[\del /\del x^\be] + \de A_\be \frac[\del /\del A_\be]$ are symmetries, just the one generated by gauge transformations on $P$, for which we have 
\eq{
\Lie_X A_\nu = \de x^\be d_\be A_\nu  -\de A_\nu = \de x^\be F_{\be \nu} + \na_\nu \xi_{V}
}
The components $(\de x^\be, \xi_{V})$ are in fact independent. We can further expand
\eqs{
\de S =&  \int_{\del D}  -u_\al ( \frac[1/4]F^{\mu\nu} F_{\mu\nu} \de_\be^\al  \de x^\be  -  F^{\al\nu} (\de x^\be F_{\be \nu} + \na_\nu \xi_{V} ))   d\ze=\cr
 =&  \int_{\del D}  u_\al \(   \( F^{\al\nu} F_{\be \nu} -\frac[1/4]F^{\mu\nu} F_{\mu\nu} \de_\be^\al \) \de x^\be   - F^{\al\nu}  \na_\nu  \xi_{V} \)   d\ze
}
Then, we have
\eqs{
&
\frac[\de S/\de x^\be] =  u_\al (F^{\al\nu}F_{\be\nu} -\frac[1/4]F^{\mu\nu} F_{\mu\nu} \de_\be^\al  ) =u_{\al}  \hat T^\al_\be  \cr
&\na_\nu \(\frac[\de S/\de A_\nu]\) = - \na_\nu \(u_\al F^{\al\nu} \)
= D_a E^a
}
As we already noticed, the Hamilton functional does not depend on $A_0$, which is then a gauge field and accounts for an extra constraint
which is associated to over-determination.

When we choose initial conditions, we need to meet the constraint $ D_a E^a=0$ on $\del D$. 
That corresponds to conservation of the electric charge, it is just a bit trivial since we are considering the vacuum theory.
On the contrary, the part associated to $\de x^\be$ corresponds to conservation of energy and momentum.

\section{ABI gravity}

We already discussed the Lagrangian for ABI model, which is 
\eq{
L=     F^i \land  L_i   +  \na k^{i} \land \( K_i-\be L_i\)     - \frac[1/2] \ep_{ijk}  k^{i}\land   k^{j}\land \(  \(\be^2-1\)  L^k -2\be K^k \) 
}
where, for the sake of simplicity, we neglect the cosmological constant and
 here we also set $\be=\ga$, as it is usually done.

The theory is first order in $A^i_\mu$ and $k^i_\mu$ only, so that we have 2 momenta
\eq{
p_i^{\mu\al} = \Frac[\de L/\del d_\al A^i_\mu] = \ep^{\al\mu\rho\si} L_{i\rho\si}
 \qquad
\pi_i^{\mu\al}=  \Frac[\de L/\del d_\al k^i_\mu] =  - \frac[1+\ga^2/\ga]  \ep^{\al\mu\rho\si} 
 u_\rho e_{k\si}
}
from which it follows that
\eq{
u_\al p_k^{\mu\al} =  \frac[1/2\ga] \ep_{kij} u_\al \ep^{\al\mu\rho\si} \ep^i_\rho \ep^j_\si
 \qquad\qquad
u_\al \pi_k^{\mu\al}=   0
}


The Poincar\'e-Cartan form reads as
\eq{
\Te_L= -\( p_i^{\mu\al} d_\al A^i_\mu +  \pi_i^{\mu\al} d_\al k^i_\mu-L\) d\si +  p_i^{\mu\al} dA^i_\mu \land d\si_\al +  \pi_i^{\mu\al} dk^i_\mu \land d\si_\al 
}
Its contraction along $X= \de x^\be \frac[\del/\del x^\be] + \de A^i_\mu \frac[\del/\del A^i_\mu]+ \de k^i_\mu \frac[\del/\del k^i_\mu]$ evaluated on the boundary is
\eqs{
X\ip\Te_L=&  \( L u_\al  \de x^\al +u_\al  p_i^{\al\mu} \Lie_X A^i_\mu + u_\al \pi_i^{\al\mu} \Lie_X k^i_\mu \) d\ze=\cr
=&  \( L u_\al  \de x^\al +u_\al  p_i^{\al\mu} \Lie_X A^i_\mu  \) d\ze
}
from which we infer
\eq{
\Frac[\de S/\de k^k_\mu] = u_\al  \pi_k^{\al\mu} = 0 
 \qquad
 \Frac[\de S/\de A^k_\mu] u_\mu=  u_\mu u_\al  p_k^{\al\mu} =0
}
The Hamilton functional depends on the tangent part of the connection, $A^i_a$.
Moreover, we have
\eqs{
 \Frac[\de S/\de A^k_a] 
 =& \frac[1/2\ga] \ep_{kij}  \ep^{abc} \ep^i_b \ep^j_c
  = \frac[1/\ga]  E^a_k  \cr
}

The Lie derivative of the connection is
\eq{
\Lie_X A^i_\nu = \de x^\be F^i_{\be \nu} + \na_\nu \xi^i_{V}
}
hence we have
\eqs{
X\ip\Te_L=&   \( ( u_\al p_i^{\al a} F^i_{\be a}+L u_\be) \de x^\be    + u_\al p_i^{\al a} D_a \xi^i_{V}  \) d\ze=\cr
   =& \( \( \frac[1/\ga]  E^a_i F^i_{\be a}+L u_\be\) \de x^\be    - \frac[1/\ga]  D_a  E^a_k  \xi^i_{V}  \) d\ze\cr
}
That gives us the {\it Gauss constraint} and, moreover, we have
\eq{
\Frac[\de S/ \de x^\be]=  \frac[1/\ga]  E^a_i F^i_{\be a}+L u_\be
}
from which we get
\eq{
\Frac[\de S/ \de x^\be] u^\be =  \frac[1/\ga]  E^a_i F^i_{\be a} u^\be -L 
\qquad
\Frac[\de S/ \de x^\be] \del_b x^\be =  \frac[1/\ga]  E^a_i F^i_{b a}
}
That gives us the second boundary equation (the momenta constraint) as well as
\eq{
\Frac[\de S/ \de x^\be] u^\be = -\frac[1/2\ep ]\( \ep^{ij}{}_{k}  F^k_{ab}   +2  ( 1+\ga^2)      \hat k^{i}_{[a}   \hat k^{j}_{b]} \) E^{a}_i E^{b}_j 
}

Accordingly, we have the constraint equations
\eq{
D_a  E^a_k =0
\qquad
E^a_i F^i_{b a} =0
\qquad
\( \ep^{ij}{}_{k}  F^k_{ab}   +2  ( 1+\ga^2)      \hat k^{i}_{[a}   \hat k^{j}_{b]} \) E^{a}_i E^{b}_j =0
}
directly from the Poincar\'e-Cartan form $\Te_L$ (or, equivalently, $\Te_H$).

\section{Conclusions and Perspectives}

Here we sketched one particular framework for covariant Hamiltonian formulation. 
There are different ones which slightly differ from this.

Our framework is based on Poincar\'e-Cartan forms, which are defined in both a Lagrangian and Hamiltonian context.
Here we introduced them for first order field theories, although they are canonical at any order in mechanics, while in higher order field theories there is one Poincar\'e-Cartan form for each connection on the base manifold.
For second order field theory, one can still choose one canonical representative, at orders higher than 2 the dependence on the connection is unavoidable.

Although non-unicity is important in a general analysis from  a mathematical physics viewpoint, here we aim to an application to a particular first order theory so that it does not play a relevant role.  
In particular, Poincar\'e-Cartan form provides us with a geometric relation with the Hamilton principal functional and, consequently, with constrain equations, 
once again without relaying on functional space, but rather algebraically on equations.

One can  summarize all we did until now by saying that we went after a way of describing the {\it physical states} of a covariant system.
All in all, a {\it state} is what uniquely determines a solution (which in a covariant setting is a {\it history} of the system, 
determining a solution means to uniquely determine the evolution of the system).
That is relevant both for a classical and quantum viewpoint.
As a result of our analysis, at least in this context, a physical state is represented by the value of fields at the boundary of a compact bubble, which {\it moreover} satisfy boundary equations, if any.
We already discussed that: from a classical viewpoint, it guarantees one can find a solution to the Cauchy problem and then to the original covariant equations.

From a quantum viewpoint, it is natural to quantize the physical state, i.e.~pre-quantum configurations obeying boundary equations, which is the quantization scheme we described.

Further  investigation have to be devoted to highlight a direct link between constraint equations and conservation laws.
The relation is very apparent in mechanics where conservation laws are related to first integrals, namely to functions on the phase space the conservation of which is, in fact,  consequence of equations of motion. It is well known that first integrals can be used to replace some of the equations of motion and, being functions of the initial conditions, namely on the phase space, they behave as constraint equations.

In field theories, conservation laws are not functions, they are {\it currents}, namely $(m-1)$-forms. The relations between conservation laws and boundary equations are well-known at the level of functionals.
Further investigation is due to consider whether a more direct link at variational level can make possible to get a covariant framework for boundary equations as it happens for conservation laws.
In particular, we have quite a well-developed framework for covariant conservation laws in a wide class of field theories (namely, gauge natural theories)
that provides covariant quantities from which each observer can eventually get its own relative quantities. 
It would not be much of a surprise the same thing happened for boundary equations. Moreover, the covariant framework for conservation laws generalizes some structures (e.g.~Bianchi identities) which also play a role with boundary equations and it may allow a general and easier approach which avoid the {\it a priori} breaking of general covariance, 
as one does in ADM framework.

Although this might be important in general, it would not change much in the specific case of LQG we are here considering.

As far as LQG is concerned, before going on to quantization, we need to better discuss connections space (and connection holonomies) and some group structures to be used later on to discuss spin networks.

\section*{Acknowledgements}

We also acknowledge the contribution of INFN (Iniziativa Specifica QGSKY and Iniziativa Specifica Euclid), the local research project {\it  Metodi Geometrici in Fisica Matematica e Applicazioni (2023)} of Dipartimento di Matematica of University of Torino (Italy). This paper is also supported by INdAM-GNFM.
We are also grateful to S.Speziale and C.Rovelli for comments.

L. Fatibene would like to acknowledge the hospitality and financial support of the Department of Applied Mathematics, University of Waterloo where part of this research was done.

\end{document}